\documentclass[12pt]{article}
\input epsf
\newcommand{\be}{\begin{equation}}
\newcommand{\ee}{\end{equation}}
\newcommand{\bea}{\begin{eqnarray}}
\newcommand{\eea}{\end{eqnarray}}
\newcommand{\ba}{\begin{array}}

\newcommand{\ea}{\end{array}}

\def\bbox{{\,
\lower0.9pt\vbox{\hrule \hbox{\vrule height 0.2 cm
\hskip 0.2 cm \vrule height 0.2 cm}\hrule}\,}}
\newcommand{\dsl}{\pa \kern-0.5em /}

\def\tr{{\rm tr}}
\def\Tr{{\rm Tr\,}}

\def\ds{\raise.15ex\hbox{/}\kern-.57em\partial}
\def\Ds{\,\raise.15ex\hbox{/}\mkern-13.5mu D}
\def\tr{{\rm tr}\,}
\def\Tr{{\rm Tr}\,}
\def\n{\nonumber}
\newcommand{\beq}{\begin{equation}}
\newcommand{\eeq}{\end{equation}} \newcommand{\beqn}{\begin{eqnarray}}
\newcommand{\eeqn}{\end{eqnarray}}

\begin{document}

\baselineskip 18pt


\begin{titlepage}
\vfill
\begin{flushright}
KIAS-P02055\\
hep-th/0209009\\
\end{flushright}

\vskip 2cm

\begin{center}
\baselineskip=17pt 
\centerline{ \Large \bf M-theory on Less Supersymmetric PP-Waves}

\vskip 10.mm
{ Kimyeong M. Lee$^\star$, }
\vskip 0.7cm
{\small\it 
School of Physics, Korea Institute for Advanced Study\\
207-43, Cheongryangri-Dong, Dongdaemun-Gu, Seoul 130-012, Korea}
\end{center}
\vskip 1.5cm
\par
\baselineskip 16pt
\begin{quote}{\small
There are fractionally supersymmetric pp-waves in 11 dimensional
supergravity. We study the corresponding supersymmetric Yang-Mills
matrix dynamics for M-theory and find its superalgebra and vacuum
equations. We show that the ground state energy of the Hamiltonian
with nontrivial dynamical superysmmetry can be zero, positive or
negative depending on parameters.  }\end{quote}

\vfill
\vskip 5mm
\hrule width 5.cm
\vskip 5mm
\begin{quote}
{\small
\noindent 
$^\star$E-mail: klee@kias.re.kr
}

\end{quote}
\end{titlepage}
\setcounter{equation}{0}

\newpage
\baselineskip 17pt

The 32 supersymmetric deformed matrix mechanics has been discovered by
Berenstein, Maldacena, and Nastase~\cite{bmn}.  This deformed matrix
theory can be regarded either as dynamics of D0 particles or as
regularized dynamics of membranes of M-theory on the maximally
supersymmetric pp-wave background~\cite{bmn,DSJvR}. This fully
supersymmetric pp-wave background~\cite{kg} was found to be the
Penrose limit of $AdS_7\times S^4 $ or $AsS_4\times
S^7$~\cite{ppwaves}. It would be interesting to find out something new
about M-theory by this approach. Some initial steps toward
understanding quantum aspects of this deformed theory has been taken
in Ref.~\cite{plefka}.

The less supersymmetric pp-wave backgrounds in M-theory have been
found and classified in Refs.~\cite{cvetic,gauntlett}. The pp-wave
background with 26 supersymmetry has been found in
Ref.~\cite{michelson}.  Including the 16 kinematical supersymmetries,
the total supersymmetries can be 16, 18, 20, 22, 26, 32 in the 11
dimensional supergravity. Some of lesser supersymmetric pp-waves in 11
dimensional supergravity can be interpreted as a Penrose limit of 
intersecting branes maybe with additional antisymmetric tensor
field. However there has been little study of the corresponding
Yang-Mills matrix quantum mechanics.

In this note we explore the less supersymmetric deformed Yang-Mills
quantum mechanics. After summerizing the old results, we explore the
superalgebra, the vacuum BPS equations for `fuzzy spheres', and the
vacuum energy in the symmetric phase. We find that the Hamiltonian
does not need to be bounded from zero and the vacuum energy of the
symmetric phase of the free abelian theory can be zero, positive or
negative depending on parameters. We conclude with some comments.

Usual supersymmetric Yang-Mills quantum mechanics with 16 dynamical
supersymmetries \cite{QM} have played a crucial role in uncovering the
11-dimensional nature of M-theory~\cite{M}.  The only known proposal
for quantum formulation of M-theory~\cite{bfss} is simply a large $N$
limit of $U(N)$ SYQM with 16 supersymmetries, which can be thought of
as large $N$ dynamics of D0 branes \cite{bound} or alternatively as a
regularized dynamics of supermembranes \cite{dWHN}. The study of the
maximally or less supersymmetric mass deformed matrix theory may leads
some insight on M-theory.

The underlying  pp-wave   in 11-dimensional supergravity is
\beqn
&& ds^2 = - 2dx^+dx^- -\sum_{i,j=1}^9 A_{ij} x^ix^j (dx^+)^2 +
\sum_{i=1}^9dx^idx^i\; , \\
&& F = dx^+ \wedge  \tilde{W} \; ,
\eeqn
where $x^\pm = (x^0\pm x^{10} )/\sqrt{2}$ and $\tilde{W} = W_{ijk}
dx^i\wedge dx^j \wedge dx^k /6$.   This pp-wave is the solution of the
11-dimensional supergravity~\cite{hull} if  the symmetric matrix $A_{ij}$
and  the antisymmetric tensor   $W_{ijk}$ satisfy
\beq
\sum_i A_{ii} = -\frac{1}{12} \sum_{ijk} W_{ijk}^2 .
\label{msquare}
\eeq

The matrix theory describes the dynamics of the bosonic degrees of
freedom $X_i$ with $i=1,...,9$ and the fermionic degrees of freedom,
16-component Majorana spinor field, $\lambda_\alpha$. Both of them
belong to the adjoint representation of $U(N)$ gauge group.  With real
and symmetric 9-dimensional gamma matrices $\gamma_i$ and
$W= \sum_{i,j,k=1}^9 W_{ijk} \gamma_{ijk}/$, the Lagrangian
for the gauged matrix theory~\cite{cvetic} is
\beqn
L&=&  \frac{1}{2} \Tr \biggl\{ \sum_i (D_0X_i)^2 +
\frac{1}{2}\sum_{ij}[X_i,X_j]^2 + i\lambda^T D_0\lambda
-\lambda^T  \gamma_i[X_i, \lambda ] \biggr. \nonumber  \\
& &  \;\;\;\; \biggl. \;\;\;\; \;\;\; - A_{ij} X_i X_j +\frac{2i}{3}
W_{ijk} X_iX_jX_k - \frac{i}{4}
\lambda^T W \lambda \biggr\} \; .
\eeqn
All quantities are  real, $\Tr$ is for the group index and the
transpose of $\lambda^T$ is for the spinor index only.  When the gauge
group is $U(1)$, the above theory becomes a free theory. The symmetric
matrix $A_{ij}$ determines bosonic mass and $W$ determines fermionic
mass at the symmetric phase $X_i=0$.  The Lagrangian has  16
kinematical supersymmetries under which
\beq
 \delta A_0=0,\;\;\; \delta X_i=0  , \;\;\;
 \delta \lambda = e^{\frac{Wt}{4}} \eta_0 
\eeq
with  time independent real spinors $\eta_0$ of 16 components.

The theory  can have a dynamical supersymmetry if for   some constant
spinor $\epsilon_0$  the theory 
is invariant under the transformation, 
\beqn
\delta A_0 &=& i\lambda^T \epsilon ,   \n \\
\delta X_j&=& i\lambda^T \gamma_j \epsilon , \n \\
\delta \lambda &=& \left( D_0 X_j  \gamma_j  -\frac{i}{2}
[X_j,X_k] \gamma_{jk} +\frac{\mu}{12} X_j (\gamma_i W+ 3W\gamma_j) 
\right) \epsilon\; 
\label{trans}
\eeqn
with 
\beq
\epsilon = e^{-\frac{Wt}{12}} \epsilon_0 . \label{time}
\eeq
The condition for the Lagrangian to be invariant under the dynamical
supersymmetry is 
\beq
\left(144 A_{jk} \gamma_j - (9W^2 \gamma_i +6W\gamma_i W+ \gamma_i
W^2)  \right) \epsilon = 0 .
\label{susy}
\eeq
This condition is identical to that for the dynamical supersymmetry in
the pp-waves~\cite{cvetic,gauntlett}. (For the discussion of deformed maximal
supersymmetric matrix theories with the matter fields, see
Ref.~\cite{piljin}. Another  study on a deformated 
supersymmetric matrix theory was done in Ref.~\cite{bonelli}. However
this work overlap with our work  little.)

To proceed, we first use the SO(9) spatial rotation to diagonalize the
symmetric matrix $A$, so that
\beq
A_{ij} = \mu_i^2 \delta_{ij}\;.
\eeq
Then we choose an `ansatz' so that antisymmetric $W$ is skew-diagonal.
 The 8 dimensional Cartan
subalgebra of SO(16) are made of mutually commuting skew-diagonal
matrices.  Any antisymmetric real 16 dimensional matrix is a linear
combination of $\gamma_{ij}$ and $\gamma_{ijk}$.  In terms of gamma
matrices, one can choose two possible nonequivalent choices of the
SO(16) Cartan subalgebra,
\beqn
&& (I)\;\;\;\gamma_{12},\gamma_{34},
\gamma_{56},\gamma_{78},\gamma_{129},\gamma_{349},
\gamma_{569},\gamma_{789},  \\
&&
(II)\;\;\;\gamma_{123},\gamma_{154}, \gamma_{167}, \gamma_{246},
\gamma_{257}, \gamma_{374},\gamma_{356}, \gamma_{89}  \; .  
\eeqn
The first seven elements of the second class is related to
octonions. (With a suitable naming, they satisfy the
algebra $\{e_i,e_j\} = -2\delta_{ij} + 2\gamma_{89}c^2_{ijk}e_k$ with
the octonionic structure constants $c_{ijk}$.)  The skew-diagonal $W$
can take two inequivalent canonical forms~\cite{cvetic,gauntlett}:
\beqn
&& W_I = m_1 \gamma_{129} + m_2 \gamma_{349} +m_3\gamma_{569} +
 m_4\gamma_{789}\; , \\
&& W_{II} = n_1
\gamma_{123}+n_2\gamma_{154}+n_3 \gamma_{167}+ n_4
\gamma_{246}+n_5\gamma_{257} +n_6 \gamma_{374}+n_7\gamma_{356}\; .
\eeqn
Two family are not exclusive of each other. When at least one of $m_i$
of the four parameter family vanishes, the resulting $W_I$ would
belong to $W_{II}$ family. In general there are 45 parameters for
$A_{ij}$ and 84 parameters for $W$. After taking out 36 parameters for
SO(9) rotation, the total nontrivial parameters are 93. Thus, we are
here focusing on rather limited ansatz of 9+4 or 9+7 parameters.

To analyze the condition (\ref{susy}), we introduce 9 hermitian
matrices
\beq
U_j = -i(W+3\gamma_j W\gamma_j)\; ,
\eeq
which commute with $W$ and are also  skew-diagonal. Then the
supersymmetric conditions (\ref{susy}) become 
\beq
 (144 \mu_j^2 - U_j^2) \epsilon = 0 
\label{susy1}
\eeq
for all $j$. As  $U_j^2$ are diagonal and the above condition becomes
easier to understand. Nontrivial supersymmetries can arise by
non-vanishing spinors satisfying the above conditions. We call such 
nontrivial spinors to be `susy spinors'. The number of dynamical
supersymmetries would be the number of linearly independent susy
spinors.

Rather than using the skew-diagonal basis, we choose complex spinor
basis for each class of $W$.  For $W_I$ we choose 16 dimensional
spinors to be $=(s_1,s_2,s_3,s_4)$ with $s_a=\pm 1$ such that they are
eigenvalues of $-i\gamma_{129}$, $-i\gamma_{349} $, $-i\gamma_{569}$,
and $-i\gamma_{789}$, respectively.  On this base $W_I$ is diagonal
and its  eigenvalues are
\beq
4 i\nu_s =   i(m_1s_1+m_2s_2+m_3s_3+m_4s_4)\; .
\label{fmass4}
\eeq
One can see from the Lagrangian that the parameter $\nu_s$ for each
$s$ is the fermion mass for the corresponding spinor field at the
symmetric vacuum $X_j=0$.  The projection operator to the eigenspinor 
is
\beq
P_s = \frac{1}{16}  (1-is_1\gamma_{129})  (1-is_2\gamma_{349})
(1-is_3\gamma_{569})  (1-is_4\gamma_{789}) \; .
\eeq
Clearly this  projection operator commutes with $W_I$. 
For  $W_I$, the diagonal $U_j$ are 
\beqn
&& U_1 = U_2 = 2(2m_1s_1-m_2s_2-m_3s_3-m_4s_4) ,\nonumber \\
&& U_3 = U_4 = 2(-m_1s_1+2m_2s_2-m_3s_3-m_4s_4 ),  \nonumber  \\
&& U_5 = U_6 = 2( -m_1s_1-m_2s_2+2m_3s_3-m_4s_4),  \nonumber  \\
&& U_7 = U_8 = 2( -m_1s_1-m_2s_2-m_3s_3+2m_4s_4 ),  \nonumber  \\
&& U_9 = 4(m_1s_1+m_2s_2+m_3s_3+m_4s_4 ) .
\eeqn

For $W_{II}$, we choose 16 dimensional spinors to be $(s_1,s_2,s_3,s_4)$
with $s_a=\pm 1$ such that they are eigenvalues of $-i\gamma_{123}$,
$-i\gamma_{154} $, $-i\gamma_{167}$, and $-i\gamma_{246}$,
respectively. On this basis the eigenvalues of  $W_{II}$ is given as
\beq
 4i \nu_s =  i(n_1 s_1 +n_2s_2 +n_3s_3+n_4s_4+n_5s_2s_3s_4
+n_6 s_1s_3s_4+n_7s_1s_2s_4),
\label{fmass7}
\eeq
Again $\nu_s$ are fermion mass for the $s$ spinor field in the
symmetric phase $X_j=0$.  The projection operator to each spinor is
\beq
P_s = \frac{1}{16}  (1-is_1\gamma_{123})  (1-is_2\gamma_{154})
(1-is_3\gamma_{167})  (1-is_4\gamma_{256}) .
\eeq
This projection operator commutes with $W_{II}$.  For the seven
parameter family $W_{II}$, the diagonal matrices $U_j$ are
\beqn
&& U_1 = 4(n_1s_1+n_2s_2+n_3s_4)-2(n_4s_4+n_5s_5+n_6s_6+n_7s_7 ), \nonumber \\
&& U_2 = 4(n_1s_1+n_4s_4+n_5s_5)-2(n_2s_2+n_3s_3+n_6s_6+n_7s_7 ), \nonumber \\
&& U_3 = 4(n_1s_1+n_6s_6+n_7s_7)-2(n_2s_2+n_3s_3+n_4s_4+n_5s_5 ),  \nonumber \\
&& U_4 = 4(n_2s_2+n_4s_4+n_6s_6)-2(n_1s_1+n_3s_3+n_5s_5+n_7s_7 ),  \nonumber \\
&& U_5 = 4(n_2s_2+n_5s_5+n_7s_7)-2(n_1s_1+n_3s_3+n_4s_4+n_6s_6 ),  \nonumber \\
&& U_6 = 4(n_3s_3+n_4s_4+n_7s_7)-2(n_1s_1+n_2s_2+n_5s_5+n_6s_6 ),  \nonumber \\
&& U_7 = 4(n_3s_3+n_5s_5+n_6s_6)-2(n_1s_1+n_2s_2+n_4s_4+n_7s_7 ),  \nonumber \\
&& U_8 = U_9= -2(n_1s_1+n_2s_2+n_3s_4 +n_4s_4+n_5s_5+n_6s_6+n_7s_7 ).  
\eeqn

For the dynamical supersymmetry to exist, the susy conditions
(\ref{susy1}) should be true for at least one spinor
$(s_1,s_2,s_3,s_4)$. Once this is true, it is also true for
$-(s_1,s_2,s_3,s_4)$ and so the dynamical supersymmetry increases by
two.  The number of such dynamical supersymmetry would be $2K$. The
total supersymmetries including the kinetic ones is $16+2K$. When
$K\ge 1$, we can choose one of such supersymmetric spinor to be
$\pm(1,1,1,1)$ without loss of generality.  The susy condition
(\ref{susy1}) fixes the mass parameters $\mu_i$ up to sign.  We choose
the sign of $\mu_i$ so that
\beq
\mu_i = \frac{1}{12} U_i(s_1=s_2=s_3=s_4=1) .
\label{bmassi}
\eeq
They are mass parameters for the bosonic $X_i$ variables in the
symmetric phase. For generic $m_i$ parameters there would be no more
spinors satisfying the conditions (\ref{susy1}), and so the total
number of supersymmetries would be $18=16+2$.  The
condition~(\ref{msquare}) for the pp-wave becomes  
\beq
\sum_i\mu_i^2 -\sum_{|s|} \nu_s^2 = 0 .
\eeq
where $|s|$ denots the spinor up to the overal sign.
This condition can  easily be shown to hold   for either four or
seven parameter families.

When these parameters are more constrained, there would be more susy
spinors satisfying (\ref{susy1}) and so there would be more
supersymmetries.  For a given number of supersymmetry $16+2K$, the
projection operators to susy spinors, $P$, would be the sum of that
for each susy spinor. As the susy spinors come in pair, $P$ is
invariant under the overall sign change of $s_i$, and so there is no
odd products of $s_i$. Thus, the $P$ is real and symmetric and so
generically
\beq
P = \frac{2K}{16} (1+ a_i \gamma_i + b_{ijkl} \gamma_{ijkl}) 
\eeq
Note that $2K= \tr P_K$ with $\tr$ being the trace over spinor index.
Also, $P$ commutes with $W$ and $U_j$'s.  The supersymmetric condition
(\ref{susy1}) becomes
\beq
U_j P = \pm  12 \mu_j P
\eeq
for all $j$. Note that in our convention, the image of $P$ always
contains $s=\pm (1,1,1,1)$.  The possible values of $K$ for the four
parameter family is $K=0,1,2,3,4,8$ and those for the seven parameter
family is $K=0,1,2,3,4,5,8$~\cite{cvetic,gauntlett,michelson}.

 As the fermion mass $\nu_s$ is proportional to the eigenvalue of $W$,
the dynamical supersymmetry for the direction along the susy spinor
$s=(s_1,s_2,s_3,s_4)$ would be time-independent if $\nu_s$ vanishes.
Many examples of four or seven parameter families of partially
supersymmetric pp-waves have been studied in
Refs.~\cite{cvetic,gauntlett}.

For the four parameter family of $W_I$, the generic supersymmetry is
18.  The case where $ W = m_1 \gamma_{129} + m_2 \gamma_{349}+
m_3\gamma_{569} $ and $\mu_i$ in (\ref{bmassi}) with $m_4=0$ has 20
supersymmetries with $P=
(1-\gamma_{1234}-\gamma_{1256}-\gamma_{3456})/4$.  If $m_3=0$, there
are 24 supersymmetries with $P = (1-\gamma_{1234})/2 $.

Among seven parameter family, one interesting case~\cite{michelson} is
such that
\beq
W= \mu \left( \gamma_{123}+\gamma_{154} +\gamma_{167}-3\gamma_{246}
+\gamma_{257}+ \gamma_{374} +\gamma_{356} \right) 
\label{26}
\eeq
with the mass parameters
\beq
\mu_1=\mu_2=...=\mu_7=  \mu , \, \, \mu_8=\mu_9 = \frac{\mu}{2} .
\eeq
In this case there are 10 susy spinors,
\beq
 \pm (1,1,1,1), \pm (-1,-1,-1,1), \pm (-1,1,1,1),  \pm (1,-1,1,1), \pm 
(1,1,-1,1) \, 
\eeq
and so  the total supersymmetries is 26. This model
is supposed to be unique modulo the usual change of coordinates.
The projection operator is the sum of that for each susy spinors
and so 
\beq
P = \frac{5-\gamma_{1234567}W}{8}.
\eeq

Let us  quantize the theory with the commutation relations 
\beq
[X_i^{ab},\pi_j^{cd}] = i\delta_{ij} \delta^{ac}\delta^{bd}, 
\;\;\; \{ \lambda^{ab}_\alpha, \lambda^{cd}_\beta \} = \delta_{\alpha\beta} 
\delta^{ac}\delta^{bd}
\eeq
with $\pi_j = \dot{X}_j$.
The supercharge  is
\beq
Q_\alpha =\left( \lambda^T \left(\pi_j\gamma_j -\frac{i}{2}
[X_j,X_k]\gamma_{jk}    +\frac{1}{12} X_j(\gamma_j W+3W \gamma_j)
\right) P\right)_\alpha .
\label{supercharge}
\eeq
The infinitesimal transformation (\ref{trans})  is induced by
\beq
\delta X_i = -[Q^T\epsilon, X_i] , \;\; \delta\lambda =
-[Q^T\epsilon,\lambda] ,
\eeq
where $\epsilon = e^{-\frac{Wt}{12}} \epsilon_0 $ with arbitrary 16
component constant spinor $\epsilon_0$. The projection operator $P$
restricts $\epsilon_0$ to  susy spinors. 

With the identity 
\beq
\lambda_\alpha \lambda_\beta = \frac{\delta_{\alpha\beta}}{2}
+\frac{(\gamma_{ij})_{\alpha\beta}}{32}  \lambda^T  \gamma_{ij}
\lambda +\frac{(\gamma_{ijk})_{\alpha\beta}}{96}
\lambda^T\gamma_{ijk}\lambda ,
\eeq
the anticommutation relation between supercharges is 
\beq
\{Q_\alpha,Q_\beta\} = 2H P_{\alpha\beta} +  L_{ij}
(R_{ij})_{\alpha\beta}  +i(P\gamma_i P)_{\alpha\beta} \Tr[ X_i {\cal
G} ] ,
\label{susya}
\eeq
where 
\beqn
R_{ij} &=& -\frac{i}{24}P (\gamma_{ij}U_j -U_j \gamma_{ji} )P
\nonumber \\ 
&=&-\frac{i}{24}P (U_i\gamma_{ij} - \gamma_{ji}U_i )P ,
\eeqn
and $L_{ij} =\Tr( X_i\pi_j-X_j\pi_i
-\frac{i}{4}\lambda^T\gamma_{ij}\lambda)$ with ${\cal G}
=i[X_i,\pi_i]+ \lambda^T\lambda$. As the Gauss law constraint is
${\cal G}=0$, the last term does not affect to  physical states.  As
$U_j P_N = \pm \frac{\mu_j}{12} P_N$, $R_{ij}$ does not vanish only if
$\mu_i^2=\mu_j^2$. Thus the rotation $L_{ij} (R_{ij})_{\alpha\beta}$
leaves the bosonic mass parameter $\mu_i$ invariant.  As the
Lagrangian is invariant under supersymmetry, there is a trivial
identity
\beq
i\gamma_{ijk} \biggl(
 P(\gamma_{ijk}U_i+U_i\gamma_{ijk})P\biggr)_{\alpha\beta} = - 8W
P_{\alpha\beta},
\eeq
which can be used to show that
\beq
[\gamma_{jk}, W] (R_{ij})_{\alpha\beta} = 0.
\label{fmasseq}
\eeq
This implies that the fermion mass parameter $\nu_s$ is also invariant
under $L_{ij} (R_{ij})_{\alpha\beta}$, and also that 
\beqn
&& [L_{ij} R_{ij}, H]=0, \\
&& [L_{ij} R_{ij} , Q^T_\alpha] = -\frac{i}{2} (Q^T
P\gamma_{ij}P)_\alpha \; .
\eeqn
Thus, $L_{ij} (R_{ij})_{\alpha\beta}$ are conserved angular momentum
for all $\alpha\beta$ They also leave the particle mass mass
invariant, as expected.  Also the
part of the superalgebra is
\beq
[H, Q_\alpha]= -\frac{i}{12}  (Q^T W)_\alpha\; ,
\eeq
and so the time dependent charge $Q(t)^T = Q^T e^{-\frac{Wt}{12}} $ is
conserved or $dQ(t)^T/dt =i [H,Q^T] + \partial Q^T/\partial t = 0$.

When  $W=m_1\gamma_{129}+m_2\gamma_{349}$ with
$P=(1-\gamma_{1234})/2$, the  angular momentum part of the
superalgebra  becomes
\beqn
L_{ij} R_{ij} &=&\biggl( \frac{(2m_1-m_2)}{3} L_{12}  +
\frac{(2m_2-m_2)}{3} L_{34} \biggr)P\gamma_9 P \nonumber \\
& & + \frac{(m_1+m_2)}{6} \;\;  \sum_{i,j=5,6,7,8} L_{ij}  P \gamma_{ij}
\gamma_{129} P \;.
\label{ang24}
\eeqn
In this case $[W, \gamma_{ij} (R_{ij})_{\alpha\beta}] = 0 $ as it
should be.

By taking the trace of Eq.~(\ref{susya}) and introducing 
\beq
Z=  - \tr  L_{ij} R_{ij}, 
\eeq
one can have a  bound on Hamiltonian,
\beq
H\ge \frac{1}{4K} Z.
\label{hbound}
\eeq
As we will see, the above bound would not be  saturated for some
ground state.

When the momentum $\pi_j$ and the fermion field are put to zero, the
angular momentum vanishes and so the Hamiltonian should be positive,
or the potential should be put into the sum of the positive terms. As
there are several Myers terms~\cite{myers}, it is not obvious how it
is done.  After inspecting the potential part of the supercharge
(\ref{supercharge}), one can see that it is natural to consider 
\beq
V=\left(-\frac{i}{2} [X_j,X_k] \gamma_{jk} + \frac{1}{12} X_j(\gamma_j W+
3W \gamma_j \right) P  ,
\nonumber
\eeq
which can be expanded as  a linear combination of products of gamma
matrices. {}From the superalgebra (\ref{susya}), we can read that the
potential energy should be
\beqn
U&\equiv&  \frac{\mu_j^2}{2} X_j^2 - \frac{i}{3} W_{ijk}X_iX_jX_k -\frac{1}{4}
[X_i,X_j]^2 \nonumber \\
&=& \frac{1}{2N} \tr \Tr V^T V .
\eeqn
The condition for the ground state would be $V=0$. Each coefficient of
the products of gamma matrices in $V$ should vanish, which become the
equations for the classical vacuum configurations. For the maximally
symmetric case the vacuum configurations are the collection of fuzzy
spheres, or giant gravitons~\cite{bmn,gg}. (For the discussion of more
general BPS configurations in the maximal case, see Ref.~\cite{pp}.)

As an example let us consider again the case $W=m_1\gamma_{129}+ m_2
\gamma_{349}$ with $P= (1-\gamma_{1234}) /2$. The above express for
the potential becomes 
\beqn
2U &=& (F_{12} +F_{34} + \frac{\mu}{3} X_9 (m_1+m_2))^2 +
(F_{13}+F_{42})^2 + (F_{14}+F_{12})^2 , \nonumber \\ 
& & + (F_{19}+ \frac{\mu}{6}
X_2(-2m_1+m_2))^2 + (F_{29}+\frac{\mu}{6} X_1 (2m_1-m_2))^2 ,\nonumber
\\
&& + 
(F_{39}+\frac{\mu}{6}X_4(-2m_2+m_1))^2+(F_{49}+\frac{\mu}{6}X_3(2m_2-m_1))^2
+ ... \; ,
\eeqn
where $F_{ij}  = -i[X_i,X_j]$ and the dots denote the sum of simple squares.
The bosonic vacuum configuration should satisfy 
\beqn
&& F_{12} +F_{34} + \frac{1}{3} X_9 (m_1+m_2) =0, \nonumber \\
&& F_{13}+F_{42}=0,\;\;\; F_{14}+F_{12} = 0, \nonumber \\ 
&& F_{19}+ \frac{1}{6} X_2(-2m_1+m_2)=0, \nonumber \\
&& F_{29}+\frac{1}{6} X_1 (2m_1-m_2) =0, \nonumber \\
&& F_{39}+\frac{1}{6}X_4(-2m_2+m_1) =0, \nonumber \\
&& F_{49}+\frac{1}{6}X_3(2m_2-m_1) =0,
\eeqn
and the rest of $X_j$ vanish.  At the moment, it is not clear what
is the general solution of the above equations. In the  case with
$m_1=m_2=m>0$, one can have a solution
\beqn
&& X_1=X_3= -\frac{m}{3\sqrt{2}} J_1, \nonumber \\
&& X_2=X_4=-\frac{m}{3\sqrt{2}}J_2, \nonumber \\
&& X_9 = -\frac{m}{6}J_3.
\eeqn
This is a fuzzy sphere in three dimension which is rotated to each of
 13 and 24 planes by 45 degrees.

The simplest quantum property is the vacuum energy at the symmetric
phase $X_i=0$. This vacuum is unique in the abelian theory which is
free. So we just focus on the abelian model with $K\ge 1$.  The
bosonic vacuum energy is
\beq
 E_{b} = \frac{1}{2} \sum_j |\mu_j |,
\eeq
where the mass parameters are given in Eq.~(\ref{bmassi}).  With our
spinor basis, there are 8 fermion degrees of freedom, one for each
$\pm(s_1,s_2,s_3,s_4)$ with mass $|\nu_s|$ of Eq.~(\ref{fmass4}) or
(\ref{fmass7}). The fermionic vacuum energy is then
\beq
E_{f} = -\frac{1}{2} \sum_{|s|} |\nu_s|.
\eeq
The question is whether the sum of the vacuum energy vanishes.  When
one of the mass parameter, say $m_1$, is very large compared with other
mass parameters $m_i, i\neq 1$, one can see that the bosonic and
fermion contributions are just given by the leading mass and their sum
vanishes for either four or seven parameter families. Thus in this limit the
supersymmetry is preserved. When the parameters are compatible, the
story becomes more complicated. 

Let us again consider the 24 supersymmetric case with
$W=m_1\gamma_{129}+ m_2\gamma_{349}$. The total vacuum energy is then 
\beq
E_0 = \frac{1}{6}\left( |2m_1-m_2|+ |2m_2-m_1| -3 |m_1 -m_2| \right).
\eeq
With normalization $m_2=1$, it is given be 
\beq
E_0 = \left\{ \begin{array}{ll} 
               0 & {\rm for} \;\;\;   m_1< 1/2 , \;\, m_1> 2 \\
               \frac{1}{3} (2m_1-1)  & {\rm for} \;\;\;  1/2< m_1 < 1 \\
               \frac{2-m_1}{3} & {\rm for} \;\;\; 1<m_1< 2 
               \end{array} \right.\; .
\eeq
One can see that the vacuum energy becomes positive for $1/2 <
\frac{m_1}{m_2} < 2$.  The term $Z$ vanishes in this case and so the
vacuum energy being nonnegative is consistent with the
bound~(\ref{hbound}). This bound for the vacuum state is not saturated
when the vacuum energy is positive. However there is a nontrivial
angular momentum part in the superalgebra as one can see from
Eq.~(\ref{ang24}).  We note that such transition of vacuum energy
occurs when either fermion or boson mass vanish. As one crosses such
boundary in the parameter space, some creation operator becomes
annihilation operator and vice versa. The bosonic part of the angular
momentum can be witten as $L\sim a^\dagger b- ab^\dagger$, with $a,b$
being annihilation operator along each spatial direction, and so its
expectation value for the ground state always vanishes. The fermionic
part of the angular momentum may develop nonzero expectation value due
to the level cross. In the parameter region where the vacuum energy is
positive the ground state of the case under discussion cannot be
singlet under supersymmetric transformation as $Z=0$ and so the
supersymmetry is spontaneously broken.  Another interesting point is
where $m_1+m_2=0$ in which case the dynamical supersymmetry becomes
time independent and the vacuum energy vanishes.

Let look at bit more complicated case of $W_I$ with nonzero
$m_1m_2m_3\neq 0$, $m_4=0$. In this case there is only 20
supersymmetries, no more or less.  The projection operator is $P =
(1-\gamma_{1234}-\gamma_{1256}-\gamma_{3456})/4$ and the central term
is nonvanishing with
\beq
Z= \frac{1}{6}(m_1+m_2+m_3) L_{78}.
\eeq
The vacuum energy of the free theory is more complicated. The three
parameter spaces is divided into many subregions. In some region, it
is zero, in some region it is positive, in some region it is
negative. When $m_1=m_2=m_3=m$, the bosonic masses are six $0$'s and
two $m/2$'s, one $m$ with $E_b=m$. The fermionic masses are six $m$'s
and two $3m$'s with $E_f= -\frac{3m}{2}$. Thus the total vacuum energy
is negative $E_0 = -\frac{m}{2}$.  

For the four parameter case $W_I$ with $m_1m_2m_3m_4\neq 0$, there are 
18 supersymmetries. The parameter space gets splited into many
regions where the vacuum energy can be zero, positive or negative. One
special subset is with $m_1+m_2+m_3+m_4=0$, in which case the
dynamical supersymmetry is time-independent. In this subset, the
vacuum energy can be positive or zero, but never negative. 

The seven parameter case is more complicated. We just note that the
case with 26 supersymmetric with  $W$ is given in Eq.~(\ref{26}), the
bosonic masses are seven $\mu$'s and two $\mu/2$'s with $E_b = 4\mu$
and the fermionic masses are five $3\mu/4$'s and three $5\mu/4$'s with
$E_f= -15\mu/2$ and so the total vacuum energy is positive $E_0 =
\mu/2$.  The $Z$ in the bound (\ref{hbound}) is $Z=10 L_{89}$.

In this work we have explored some basic features of the less
supersymmetric matrix model on pp-wave background. We found the
superalgebra, the classical vacuum equation, the quantum vacuum energy
at the symmetric phase. The most interesting part is that the vacuum
structure can be rather rich, leading to the spontaneously broken
supersymmetry. Notice that all this structure is there even for free
abelian theory. Thus its cause should be purely algebraic. The detail
of the vacuum structure of the less supersymmetric Yang-Mills matrix
mechanics is currently under study.

We have seen that the number of dynamical supersymmetries increases by
even integer. It may be due to our special ansatz for $A$ and $W$
tensor. It would be interesting to find the analysis for the general
case.

As argued in Ref.~\cite{cvetic,gauntlett}, the pp-wave background for
our model can arise from the Penrose limit of intersecting branes
maybe with additional background antisymmetric tensor field. It would
be interesting to figure out the implication of our work on the
pp-wave itself in M-theory.

\vskip 1cm

\noindent {\large \bf Acknowledgement} 

This work is supported in part by KOSEF 1998
interdisciplinary research grant 98-07-02-07-01-5 (K.L.). I am
grateful to D. Bak, J.-H. Park and P. Yi for many useful comments and
discussions.

\end{document}